\documentclass[%
reprint,
 amsmath,amssymb,
 aps, prl
]{revtex4-1}

\usepackage{hyperref}

\usepackage{subfigure}
\usepackage{graphicx}
\usepackage{epstopdf}

\usepackage{dcolumn}

\usepackage{etoolbox}
\usepackage{bm}

\begin{document}


\title{Universal gates based on targeted phase shifts in a 3D neutral atom array}	

\author{Yang Wang}

\author{Aishwarya Kumar}%

\author{Tsung-Yao Wu}%
\author{David S. Weiss}%

\email{dsweiss@phys.psu.edu}
\affiliation {Department of Physics, the Pennsylvania State University}

{%
}%

\date{\today}

\begin{abstract}

Although the quality of quantum bits (qubits) and quantum gates has been steadily improving, the available quantity of qubits has increased quite slowly. To address this important issue in quantum computing, we have demonstrated arbitrary single qubit gates based on targeted phase shifts, an approach that can be applied to atom, ion or other atom-like systems. These gates are highly insensitive to addressing beam imperfections and have little crosstalk, allowing for a dramatic scaling up of qubit number. We have performed gates in series on 48 individually targeted sites in a 40\% full $5\times 5\times 5$ 3D array created by an optical lattice. Using randomized benchmarking, we demonstrate an average gate fidelity of 0.9962(16), with an average crosstalk fidelity of 0.9979(2).

\begin{description}
\item[PACS numbers]

\end{description}
\end{abstract}
\maketitle



The performance of isolated quantum gates has recently been improved for several types of qubits, including trapped ions \cite{Lucas2014,Wunderlich2014,WinelandSPW2014}, Josephson junctions \cite{Martinis2014}, quantum dots \cite{Dzurak2014}, and neutral atoms \cite{SaffmanRandom2015}. Single qubit gate errors now approach or, in the case of ions, surpass the commonly accepted error-threshold \cite{knillquantum2005,ErrCorrTheory} (error per gate $ < 10^{-4}$), for fault-tolerant quantum computation  \cite{WinelandEC2004,Hennrich2011,corcolesdemonstration2015,Rainer2008}. It remains a challenge in all these systems to execute targeted gates on many qubits with fidelities comparable to those for isolated qubits \cite{divincenzo2000,laddquantum2010}. Neutral atom and ion experiments have to date demonstrated the most qubits in the same system, 50 and 18 respectively \cite{wangcoherent2015,senko2014}. The highest fidelity gates in these systems are based on microwave transitions, but addressing schemes typically depend on either addressing light beams \cite{weissanother2004,valaperfect2005,weitenbergsingle-spin2011,SaffmanRandom2015,wangcoherent2015} which are difficult to make as stable as microwaves, or magnetic field gradients \cite{Wunderlich2014,Hensinger2015} which limit the number of addressed qubits. In this report, we present a  way to induce phase shifts on atoms at targeted sites in a $5\times 5 \times 5$ optical lattice that is highly insensitive to addressing laser beam fluctuations. We further show how to convert targeted phase shifts into arbitrary single qubit gates. These high fidelity gates are only sensitive to laser fluctuations at second order in intensity and fourth order in beam pointing. We demonstrate average single gate errors across our array that are below 0.004, and present a path towards reaching the fault-tolerant threshold.

In previous work \cite{wangcoherent2015} we performed single site addressing in a 3D lattice using crossed laser beams to selectively ac Stark shift target atoms, and microwaves to temporarily map quantum states from a field insensitive storage basis to the Stark-shifted computational basis. While we use most of the same physical elements here, the crucial difference is that these new gates are based on phase shifts in the storage basis, and do not require transitions out of it. Non-resonant microwaves are applied that give opposite-sign ac Zeeman shifts for different atoms. A specific series of non-resonant pulses and global $\pi$-pulses on the qubit transition gives a zero net phase shift for non-target atoms and a controllable net phase shift for target atoms. The resultant gate fidelity is much better than our previous gate because of this gate's extreme insensitivity to the addressing beam alignment and power, the insensitivity of the storage basis to magnetic fields and vector light shifts, and the independence on the phase of the non-resonant microwave pulses. The principle of this phase gate, where optical addressing does not compromise fidelity, can be adapted to other atom and ion qubit geometries.

\begin{figure}[!h]
\includegraphics[width=8cm]{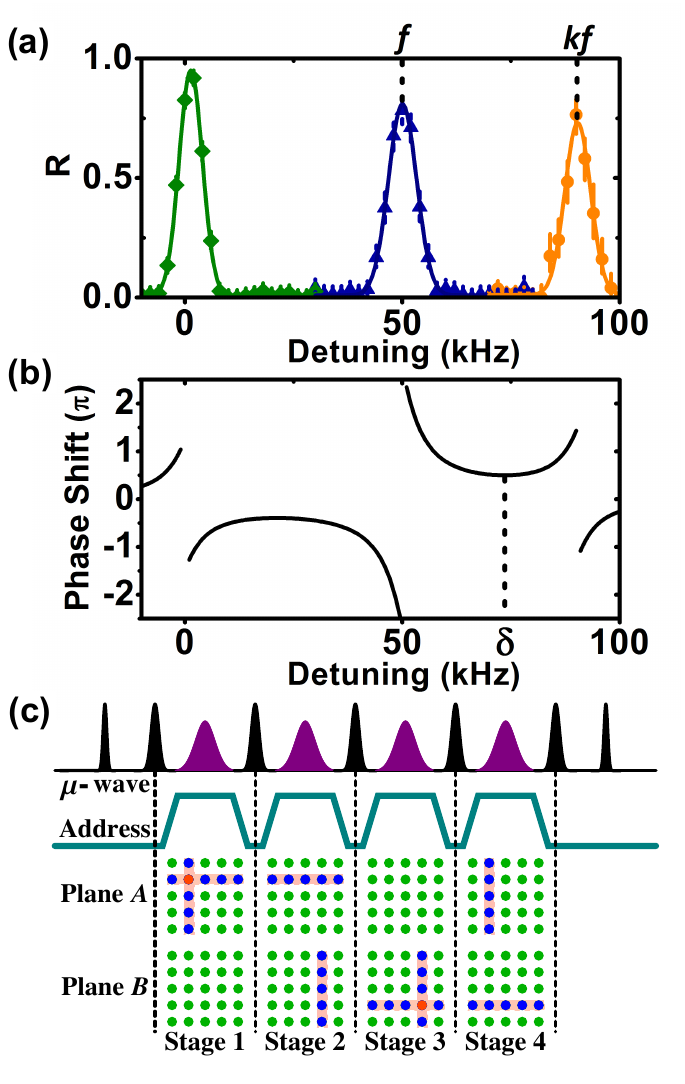}
\caption{\label{fig1}
Addressing spectrum, phase shift and timing sequence.
(a) Spectra of the $|4,0\rangle \rightarrow |3,1\rangle$ transition. We plot the ratio of the number of detected $F=3$ atoms to the initial number of atoms as a function of the detuning from the unshifted resonance. The green diamonds are due to spectators, the blue triangles are due to line atoms and the orange circles are due to cross atoms. The solid lines are fits to Gaussians. The resonance frequencies of line atoms and cross atoms are marked by $f$ and $kf$.
(b) Exact calculation of the cumulative phase shift ($\phi_{\textrm{Target}}$) on a target atom as a function of the microwave detuning from the unshifted resonance. The optimum operation point is marked by $\delta$.
(c) Addressing pulse sequence. First row: Blackman-profiled microwave pulse sequence. The black pulses (80 $\mu$s) are resonant on the $|0\rangle$ to $|1\rangle$ transition and affect all atoms. The purple pulses (120 $\mu$s) are addressing pulses detuned from the $|4(3),0\rangle$ to $|3(4),1\rangle$ transitions by $\delta$.  Second row: the addressing light intensity. The addressing light barely affects the trapping potential, so a linear ramp is optimal. It is effectively at full power for 252 $\mu$s. Third and fourth rows: schematic of addressing beams in two planes. The atom color codes are as in (a).
}
\end{figure}

Detailed descriptions of our apparatus can be found in refs 15, 21 and 22. 
We optically trap and reliably image neutral $^{133}$Cs atoms in a 5 $\mathrm{\mu m}$ spaced cubic optical lattice. The atoms are cooled to $\sim70\%$ ground vibrational state occupancy and then microwave transferred into the qubit basis, the $6S_{1/2}$,  $|3,0\rangle$ and $|4,0\rangle$ hyperfine sublevels, which we will call $|1\rangle$ and $|0\rangle$, respectively. Lattice light spontaneous emission is the largest source of decoherence, with a 7 s coherence time ($T_2^{'}$) that is much longer than the typical microwave pulse time of 80 $\mathrm{\mu s}$. We detect the qubit states by clearing atoms in the $|4,0\rangle$ states and imaging the $|3,0\rangle$ atoms that remain.

To target an atom, we cross two circularly polarized, 880.250 nm addressing beams (beam waist $\sim$ 2.7 $\mathrm{\mu m}$, Rayleigh range $\sim$ 26 $\mathrm{\mu m}$). The addressing beams can be directed to a new target in $<$ 5 $\mathrm{\mu s}$ using micro-electro-mechanical-system (MEMS) mirrors \cite{kim}. The addressing beams ac Stark shift the $m_F\neq 0$ levels of the atom at their intersection by about twice as much as any other atom. This is illustrated in Fig. 1(a), for which atoms are prepared in $|4,0\rangle$ and a microwave near the $|3,1\rangle$ transition is scanned. The resonances are visible for atoms at the intersection (orange, termed "cross" atoms), atoms in one addressing beam path (blue, termed "line" atoms), and the rest of the atoms (green, termed "spectator" atoms). The ac Stark shift for the line atoms, $f$, is chosen so that there is a region between the blue and orange peaks in which only a small fraction ($ 2\times 10^{-4}$) of atoms in any class (cross, line or spectator) makes the transition. When a microwave pulse is applied in that frequency range, atoms experience different ac Zeeman shifts depending on their class.

The addressing pulse sequence for a pair of target atoms in two planes, shown in Fig 1. (c), consists of four stages  \cite{wangcoherent2015}. The qubit-resonant spin-echo pulses (the black pulses on the microwave line in Fig 1. (c)) reverse the sign of the phase shifts, so whatever phase shifts (ac Zeeman or ac Stark) a non-target atom gets during the cross stages (the first and third) are exactly canceled by the shifts it gets during the dummy stages (the second and fourth), where there is no cross atom. In contrast, the first (second) target atom spends stage 1 (3) as a cross, stage 3 (1) as a spectator, and stages 2 and 4 (2 and 4) as a line atom. When the microwave frequency is chosen to be between the line and cross resonances, the change in the target atom's status from cross to line changes the sign of the ac Zeeman shift. Away from the resonances, the net phase shift for the target atoms is:
\begin{equation}
\label{phase}
\phi_{\textrm{Target}}\approx\frac{\Omega^2 T }{\delta-kf}-\frac{\Omega^2 T }{\delta-f}+\frac{\Omega^2 T }{\delta}-\frac{\Omega^2 T }{\delta-f}
\end{equation}
where $\delta$ is the microwave detuning from the spectator resonance, $k$ is the shift of cross atoms in units of $f$ (see Supplementary Materials),
$\Omega$ is the microwave Rabi frequency and $T$ is the pulse duration. Successive terms correspond to the integrated ac Zeeman shift on the first target atom during successive stages. The overall phase shift can be directly controlled by changing the power of the microwave field. The black curve in Fig. 1(b) shows the result of an exact calculation of the target atom's phase shift as a function of $\delta$. The phase shift minimum at $\delta =74.9$ kHz is the preferred operating point for the gate, since the shift then quadratically depends on the change in $\delta$ and thus also on the addressing beams ac Stark shift, with the coefficient of 21 $ \textrm{rad}/(\Delta f/f)^2$. For example, a 2$\%$ change in $f$ gives a 8 mrad phase shift, which in turn leads to only a $1\times10^{-4}$ gate error.  Since the intensity changes quadratically with beam alignment, the gate is sensitive to beam pointing only at fourth order.

The phase shift on target sites amounts to a rotation about the $z-$axis (an $R_z(\theta)$ gate), but a universal gate requires arbitrary rotations about any arbitrary axis. We can make an $R_y(\theta)$ gate by combining the $R_z(\theta)$ gate with global $R_x(\pm\frac{\pi}{2})$ rotations:
\begin{equation}
\label{universal}
R_y(\theta)=R_x(\frac{\pi}{2})R_z(\theta)R_x(- \frac{\pi}{2})
\end{equation}
For non-target atoms, which see the global microwave pulses but experience no $R_z(\theta)$, $R_x(\frac{\pi}{2})R_x(- \frac{\pi}{2})$ clearly has no net effect. It is straightforward to generalize this formula to obtain arbitrary rotations on a Bloch sphere for target atoms. The corresponding complete set of single qubit gates on target atoms all leave the non-target atoms unchanged.

\begin{figure}[!h]
\includegraphics[width=9cm]{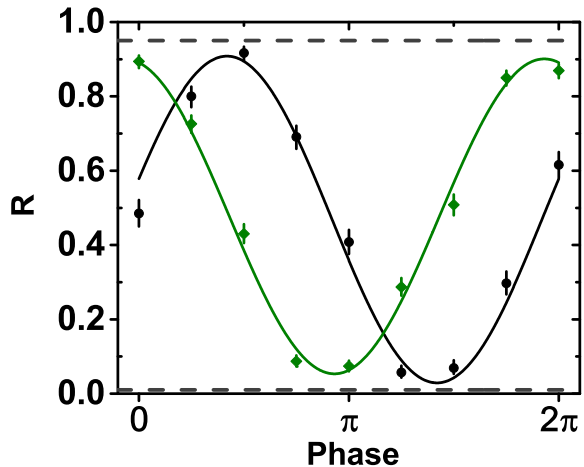}
\caption{\label{fig2}
Interference fringe of one $R_z(\frac{\pi}{2})$ phase gate applied to 48 randomly chosen sites. The black dots correspond to the target atoms, and the green dots correspond to non-target atoms. Solid lines are fits to data. Dashed lines mark the maximum and minimum possible populations. After all these gates, the net $\mathcal{F}^2 $ for target atoms is $0.94(3)$, and for the non-target atoms it is $0.94(2)$ . The average error per gate for both target and non-target atoms is $13(7) \times 10^{-4}$.
}
\end{figure}

We have demonstrated one $R_z(\pi/2)$ gate on 48 randomly chosen sites (in 24 gate pairs) within a $5\times5\times5$ array. Given the average initial site occupancy of $\sim 40\%$, an average of 20 qubits experience the phase gate during each implementation. We probe the coherence by closing the spin echo sequence with a global $\frac{\pi}{2}$ pulse whose phase we scan, as shown in Fig. 2. The green points are due to non-target atoms, and the black points are due to atoms at the 48 target sites. The corresponding curves are sinusoidal fits to the data. The dashed lines mark the maximum and minimum populations one expects given perfect gate fidelity, $\mathcal{F}^2$, defined as the square of the projection of the measured state onto the intended state, in the face of state preparation and measurement (SPAM) errors (see Supplementary Materials). From these curves we determine that the error per gate pair, $\mathcal{E}_2$, is $25(13) \times 10^{-4}$ for both target and non-target atoms on average. At least for this particular gate, most of the error is clearly common  to target and non-target atoms. The good contrast of the target atom fringe illustrates the homogeneity of the phase shifts across the ensemble. The error per gate, $\mathcal{E}$, is thus $13(7) \times 10^{-4} $ on average.  

To graphically illustrate the flexibility of our gates, we have performed an $R_z(\pi)$ rotation on a 32 site pattern (see Fig. 3). We rotate the superposition of target atoms from $\frac{1}{\sqrt{2}}(|0\rangle+|1\rangle)$ to $\frac{1}{\sqrt{2}}(|0\rangle-|1\rangle)$, and then use a $\pi/2$ detection pulse with phase of $\pi$ to return spectator atoms to $|0\rangle$ and target atoms to $|1\rangle$. Fig. 3 shows images from 5 planes summed over 50 implementations.

\begin{figure}{!h}
\includegraphics[width=9cm]{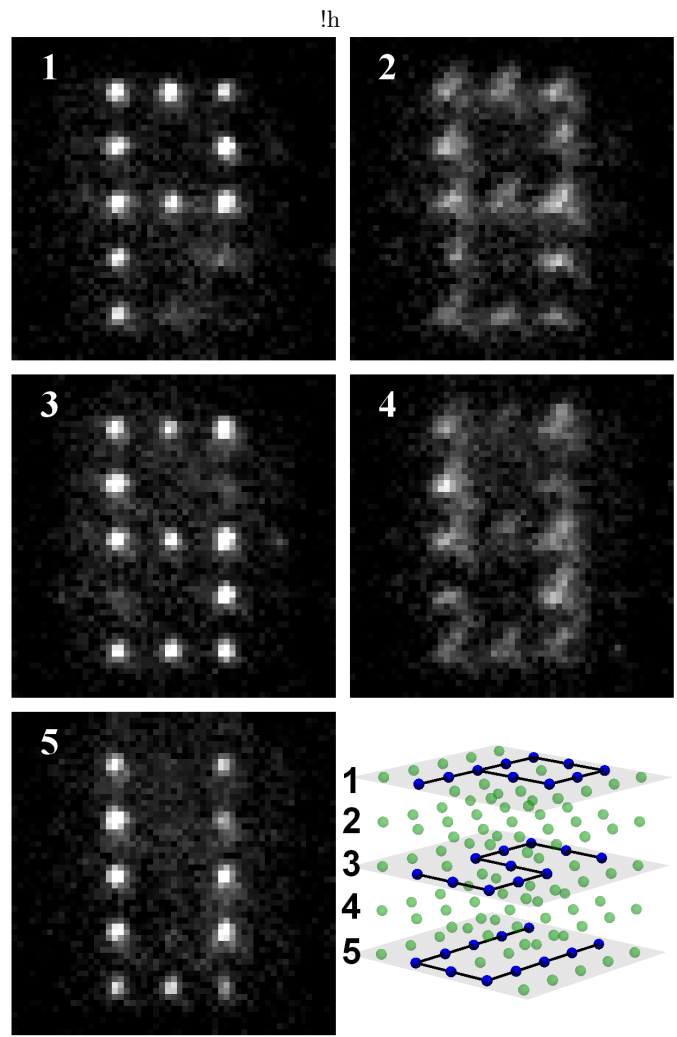}
\caption{\label{fig3} 
$R_z(\pi)$ gates performed in a specific pattern.
(1-5): Fluorescent images of 5 successive planes. Each image is the sum of 50 implementations. For clarity, the contrast is enhanced using the same set of dark/bright thresholds on all pictures to account for the shot noise. Note that there are no targeted sites in planes 2 and 4; the light collected there is from atoms in the adjacent planes. The drawing to the lower right marks the targeted sites in blue, with plane shading and blue lines to guide the eye.}
\end{figure}

To confirm the robustness of this gate, we applied an $R_z(\pi/2)$ gate to 24 targets, and measured the probability that atoms return to $|1\rangle$ after a $\pi/2$ pulse as a function of the fractional change in addressing beam intensity (see the Fig. 4 inset and Supplementary Materials). The data confirms the theoretical prediction (solid line) of extreme insensitivity to addressing beam intensity (see Fig. 1(b)).

\begin{figure}[!h]
\includegraphics[width=9cm]{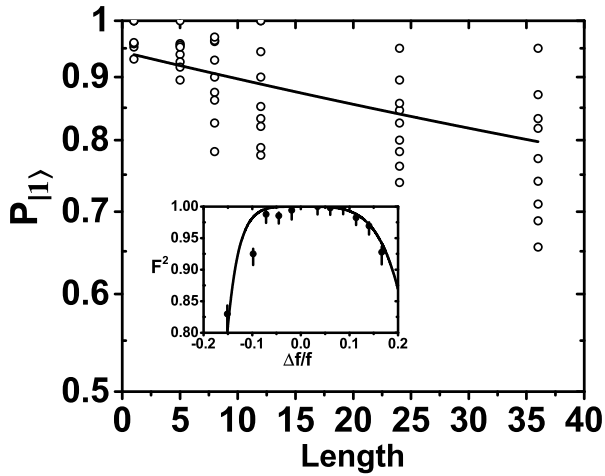}
\caption{\label{fig4}
Semi-log plot of fidelity from a randomized benchmarking sequence. We plot the probability of returning the two target atoms to $|1\rangle$ after a randomized benchmarking sequence versus the sequence length.  At each length we employ 3 randomized CG sequences, each of which is combined with 3 sets of randomized PG sequences, for a total of 9 points. Each data point is averaged over $\sim$100 implementations. A least squares fit using the $d_{\textrm{if}}=0.1128(68)$ determined from the auxiliary RB measurement on non-target atoms gives $\mathcal{E}_{2t}$ = $(55 \pm 16) \times10^{-4} $.
Inset: Fidelity of an $R_z(\pi/2)$ gate as a function of the fractional change of the addressing ac Stark shift. The points are experimental data and the curve is from the exact calculation.
}
\end{figure}

To fully characterize gate performance we employ the standard randomized benchmarking (RB) protocol \cite{knillrandomized2008}, which has been used in nuclear magnetic resonance \cite{ryanrandomized2009}, quantum dots \cite{Dzurak2014}, ion systems \cite{Wineland10em42011} and neutral atoms systems \cite{PortoNJP2010,leerobust2013,SaffmanRandom2015}. We implement RB by choosing computation gates (CG) at random from \{$R_x(\pm\frac{\pi}{2})$, $R_y(\pm\frac{\pi}{2})$\} and Pauli gates (PG) at random from \{$R_x(\pm \pi),R_y(\pm \pi),R_z(\pm \pi), I $\} where $I$ is the identity \cite{KitaevClifford2005}. Following each randomized sequence, a detection gate is applied that in the absence of errors would return either the target qubits or the non-target qubits to $|1\rangle$.  For the targets, the average value of $\mathcal{F}^2$ decays exponentially with the length of the randomized sequence:
\begin{equation}
\label{RB}
\overline{\mathcal{F}^2}=\frac{1}{2}+\frac{1}{2}(1-d_{\textrm{if}})(1-2\mathcal{E}_{2t})^l
\end{equation}
where $d_{\textrm{if}}/2$ is the SPAM error and $l$ is the number of CG-PG operations applied to the pair of target atoms.  A similar expression yields $\mathcal{E}_{2s}$ and $\mathcal{E}_{2l}$ for the spectator and line atoms respectively, the unwanted changes wrought on the non-target atoms when a random series of gates is executed on the targets. In this way we can characterize crosstalk and extract errors per gate from errors per gate pair.  For target atoms, $\mathcal{E}_t=\mathcal{E}_{2t}-\mathcal{E}_s$,  for spectator atoms, $\mathcal{E}_s=\mathcal{E}_{2s}/2$, and for line atoms, $\mathcal{E}_l = \mathcal{E}_{2l}-\mathcal{E}_s$.

First analyzing the non-target data (see Supplementary Fig. 3), we determine that $d_{\textrm{if}}=1.1(1)\times 10^{-2}$, $\mathcal{E}_s=17(2)\times 10^{-4}$, and $\mathcal{E}_l=46(9) \times 10^{-4}$.  The crosstalk, defined as the average error per gate for non-target atoms, is $ \frac{107}{123}\mathcal{E}_s+\frac{16}{123}\mathcal{E}_l =21(2) \times 10^{-4}$. Only global microwave pulse imperfections contribute to $\mathcal{E}_s$.  Their contribution to $\mathcal{E}_l$ may differ since the microwave sequence is optimized for spectators. Spontaneous emission from the addressing beams adds to $\mathcal{E}_l$, with a calculable contribution of $8\times 10^{-4}$. Fig. 4 shows the RB data for the two target atoms. Using the $d_{\textrm{if}}$ from the non-targets, the fit of Eq. 3 to the data yields $\mathcal{E}_t=38(16)\times 10 ^{-4}$. 

A summary of the errors for the target and non-target atoms is shown in Table 1. The errors can mostly be traced to microwave power stability.  We expect that reaching the state of the art \cite{Wineland10em42011,Wunderlich2014}, and perhaps tweaking our spin echo infrastructure \cite{DDReview2012} (see Supplementary Fig. 1), can bring it below $10^{-4}$. The next largest contribution is from the readily calculable spontaneous emission of addressing light. Its contribution is invariant with gate times, since $f$ must be changed inversely with gate time. Addressing with light between the D1 and D2 lines, as we do, locally minimizes spontaneous emission, but an eight-fold reduction could be obtained by doubling the wavelength, which would not dramatically compromise site addressing. The required powerful addressing beams would exert a significant force on line atoms, but deeper lattices and adiabatic addressing beam turn-on would make the associated heating negligible. In our current experiment we need to wait 70 $\mathrm{\mu s}$ after the addressing beams are turned on for our intensity lock to settle. Technical improvement there could halve the time the light is on, ultimately leaving the spontaneous emission contribution to the error just below $10^{-4}$ per gate.  That limit is based on addressing with a differential light shift. It would disappear were the same basic scheme to be employed on a 3-level system, where only one of the qubit states is strongly ac Stark shifted by the addressing light. In that case the phase shifting microwaves (or light) would simply be off-resonant from the qubit transition, and the error due to spontaneous emission would decrease proportional to the detuning.  Since the dominant target errors have the same sources as the non-target errors, improvements to the gate will correspondingly improve the crosstalk. Adapting this method to more common 1D and 2D geometries is straightforward. Only a single addressing beam is needed, and the dummy stages would be executed with half-intensity addressing beams. The same insensitivity to addressing beam power and alignment would follow.

In summary, we report on a universal single qubit gate based on targeted phase shifts. The gate is robust to addressing light fluctuations, and can be applied to large arrays of qubits, including our demanding 3D lattice geometry. Scalable addressing is an important step towards a scalable quantum computer.


\begin{table*}[!t]
  \centering
  \label{Tb1}

    \begin{tabular}{c |c | c |c  | c}
   \hline 
     Index   &   Items  & Spectator   & Line & Target  \\
    \hline \hline
    & $\boldsymbol{\mathcal{E}_2}$, error per gate pair             						& 34(4)  & 63(9)  &  55(16)  \\
      \hline
    &    $\boldsymbol{\mathcal{E}}$, error per gate  	 					             &  17(2)  & 46(9)  & 38(16) \\
   \hline
   i	&	Spontaneous emission of addressing light 	& 	 --  		& 8 	& 16 \\
  ii	& 	Phase error from fluctuations in $f$ 	  		& 	 -- 		&   --  	&0.7 \\
iii	&	Phase error from microwave fluctuations		& 	  --  		& --  	& 0.2 \\
iv	&	Off-resonant microwave transitions 			& 	--		 & 3.6 	 & 5.4 \\ 
v	&	Other sources 		   					       	& 17	 	& 	34 & 16	 \\	
   \hline
    \hline
    \end{tabular}
      \caption{ 
      Summary of the various error sources that contribute to gate errors in units of $10^{-4}$. 
      (i) The spontaneous emission rate is directly calculable from the vector light shift: $\Gamma_{sc}=3.1\times 10^{-2}$ $\textrm{s}^{-1}/$kHz.
      (ii) Addressing-induced inhomogeneous broadening (Fig. 1 (a), broadened blue and orange peaks) implies $\sim$ 2$\%$ spread of the ac Stark shift $f$, leading to an 8.4 mrad phase rotation spread and a 0.7 $\times 10^{-4}$ fidelity error.
      (iii)  Our passively stabilized microwave source limits the fidelity of individual microwave pulses (see Supplementary Materials). The phase error due to the ac Zeeman shifting microwave pulses is $\sim$ 5 mrad, leading to a $0.2 \times 10^{-4}$ fidelity error.          
      (iv) Calculated off-resonant excitation due to the phase shifting microwave pulse. Increasing $f$ by 12\% would increase spontaneous emission by 12\% but decrease this error below $1\times10^{-4}$.
       (v) Other sources of error that are neither calculated nor directly measured are presumably dominated by global microwave error. 
       Global microwave error in the spin echo infrastructure results from microwave power fluctuations and inhomogeneous broadening (see Supplementary Materials). It is essentially the only term that contributes to $\mathcal{E}_s$. 
      }
\end{table*}



\section{Acknowledgment}

This work was supported by the US National Science Foundation Grant PHY-1520976.


\makeatletter
\apptocmd{\thebibliography}{\global\c@NAT@ctr 30\relax}{}{}
\makeatother

\clearpage
\section*{Supplementary Materials}

\section{Phase gate details}
Inspection of Fig. 1 (b) shows that there are two phase shift extrema as a function of $\delta$.  The one we do not use, between the spectator and line resonances, is less sensitive to detuning, but it would increase the rate of unintentional transfer out of the qubit basis.  This is partly because there are many more spectator atoms than targets; the small amount of off-resonant transfer would get multiplied by a much larger number.  In addition, there can be atoms that see a fraction of the intensity that most line atoms see, due to misalignment of beams relative to the lattice or beam profile imperfections; this can put their resonances closer to the phase shifting microwaves.

To ensure operation near the phase shift extremum, we find it necessary to adjust the intensities of the addressing beam in a target specific way. This corrects for beam profile distortion by the MEMS mirror steering system and for the finite Rayleigh range of the addressing beams near their foci.

The deviation of the cross atom resonance, 1.8$f$, from 2$f$  (see Fig. 1 (a)) occurs because the Zeeman shift due to the bias B-field is not very much larger than the addressing beams' vector light shifts. Thus when there is only one addressing beam, the quantization axis tilts slightly toward its propagation direction. 

\section{Randomized Benchmarking}

\renewcommand{\figurename}{Supplementary Figure}
\setcounter{figure}{0}  
\begin{figure}[!b]
\centering
\includegraphics[scale=0.5]{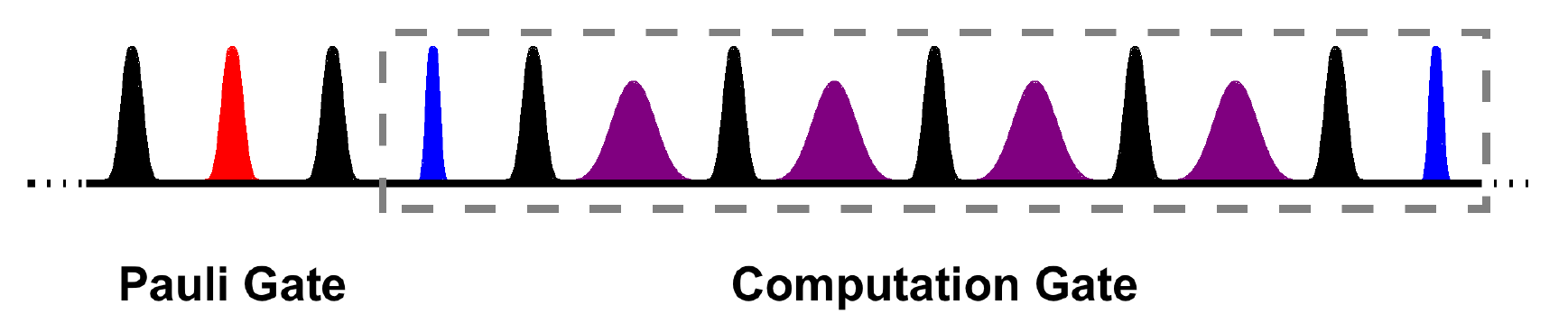}
\caption{\label{EDfig1} 
Our RB building block. These are the microwave pulses corresponding to a PG/CG pair. The red pulse is the PG. The CG  is inside the dashed rectangle: the blue pulses are $\pi/2$, the black pulses form the spin-echo type structure and the purple pulses are off-resonant microwaves that induce phase shifts. All the pulses have a Blackman profile.}
\end{figure}

We choose randomized benchmarking (RB) to characterize the performance of our gate instead of quantum state tomography for several reasons: it can distinguish gate errors from SPAM errors;  it has less resource overhead and is easily scaled to multi-qubit systems; and it reflects the fidelity of a group of operations that is not biased by a particular kind of gate.  

The basic building block of an RB sequence is a  PG-CG pair executed on two target atoms. A CG is a $\pm \frac{\pi}{2}$ rotation about the $x-$ or $y-$axis on the Bloch sphere. It contains one phase gate $R_z(\theta)$ sandwiched by two global $R_{x,y}(\pm\frac{\pi}{2})$ rotations.  A PG is implemented as a global $\pm \pi$ rotation (about the $x-$ or  $y-$axis) or the absence of one (for the Identity or a rotation about $z$). 
For the $R_z(\pm \pi)$ rotation, we frame-shift all successive microwave pulses.
We extend the spin-echo type structure to include the Pauli pulses, as shown in Supplementary Fig. 1. The $\pi/2$ and the Pauli pulses are at the rephasing points in this spin-echo type structure. While the main purpose of the echo $\pi$ pulses is to cancel the crosstalk due to addressing beams and microwave pulses, they also serve to preserve the qubit in the face of inhomogeneous broadening. Moreover, their phases can be adapted to mitigate microwave power errors, 
as described in the next section. 

We have used separate measurements to isolate the factors that contribute to the SPAM error: collisions with background gas atoms (3$\%$), imperfect state transfer from $|4,4\rangle$ to $|4,0\rangle$ (2$\%$), and  imperfect clearing ($<0.5\%$). The sum of these expected terms matches the measured $d_{\textrm{if}}$ from RB.

\section{Microwave pulse errors}

Generally, there are two types of microwave transition imperfections:  power and frequency errors.  Our passively stabilized microwave source limits our power stability. We effectively measure it by driving atoms with up to 65 $\pi$-pulses and find the average fractional error to be $\sim 3\times 10^{-3}$. The source of frequency errors in our system is the inhomogeneous broadening of the ensemble of atoms ($\sim$130 Hz) due to vibrational excitation. Although, this is too small to affect the fidelity of individual microwave pulses at the $10^{-4}$ level, state evolution between pulses affects our error cancellation scheme. Various techniques have been developed in NMR for dealing with such errors, including composite pulses \cite{bb1corpse}, and dynamical decoupling\cite{callaghan_principles_1993}. Composite pulses from the BB1 family can be made insensitive to imperfect pulse amplitude and pulses from the CORPSE families can be made insensitive to frequency errors. Since neither does both, and they come at the cost of considerably longer pulse durations, they are not suitable for our purpose. We instead implement phase cycling schemes in which the errors introduced by microwave pulses are cancelled by subsequent pulses. 

Before tackling the more complicated case of a random sequence of Clifford and Pauli gates, let us consider a simple spin echo sequence,  a repeating sub-cycle of torque vector directions given by : \{$y,y,-y,-y$\} on the Bloch sphere. Dephasing from any initial state is controlled, unlike with simpler schemes or the widely used XY class of pulses \cite{gullion_elimination_1991, souza_robust_2011}, which don't have a perfect cancellation in the absence of inhomogeneous broadening. Inhomogeneous broadening compromises error cancellation, but phase cycling still greatly improves the performance. 

In Supplementary Fig. 2 we demonstrate the performance of this pulse scheme. Supplementary Fig. 2 (a) plots the spin echo fringe with 100 $\pi$ pulses with phase cycling, with the dashed lines marking the maximum population. The almost lossless fringe contrast confirms the high fidelity of this spin-echo type structure. Supplementary Fig. 2 (b) plots this contrast versus microwave Rabi frequency. It is clear that this pulse scheme is quite robust against power fluctuation. 

\begin{figure}[!b]
\centering
\subfigure[t][]{
\includegraphics[width=3.1 in]{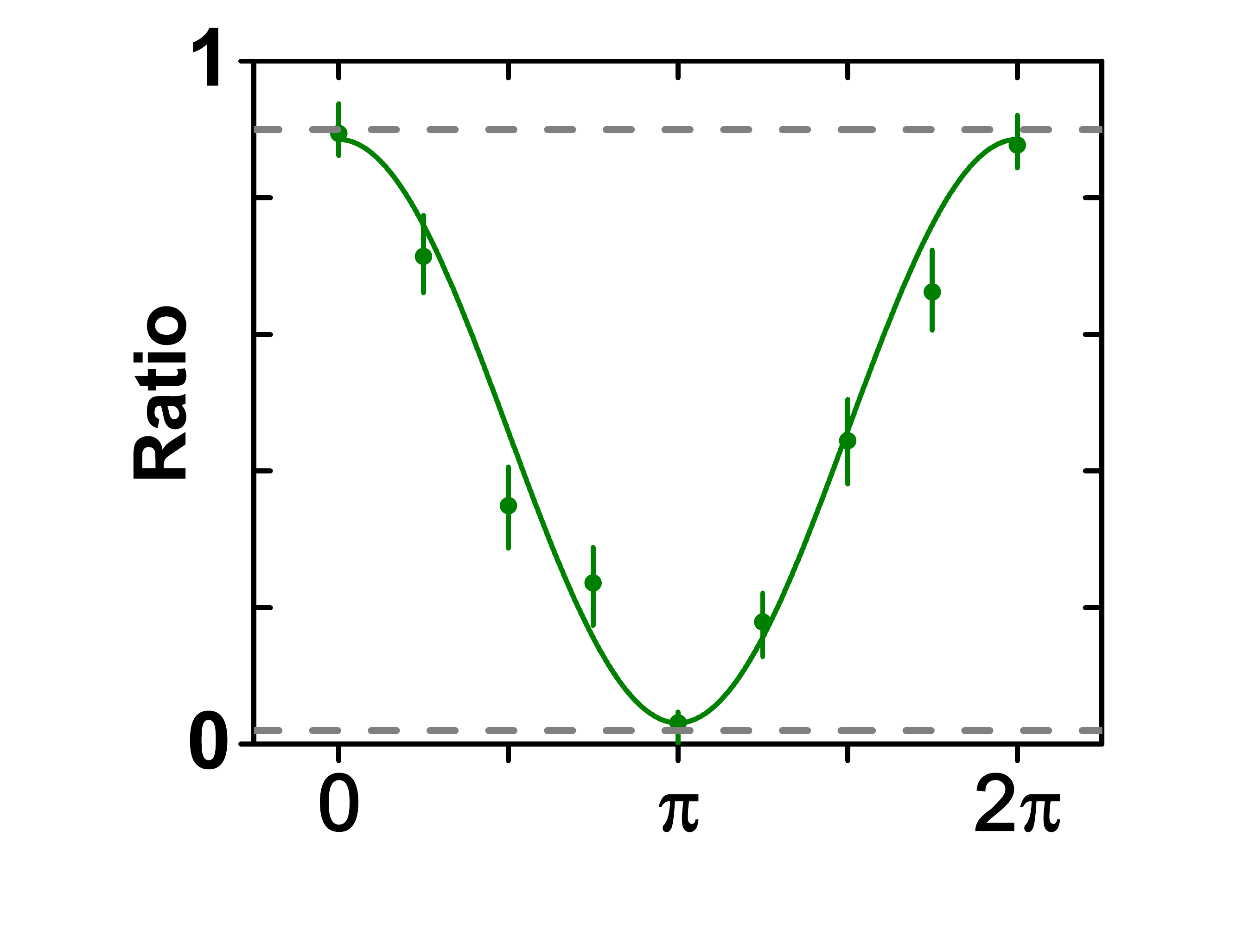}}
\subfigure[t][]{
\includegraphics[width = 3.1 in]{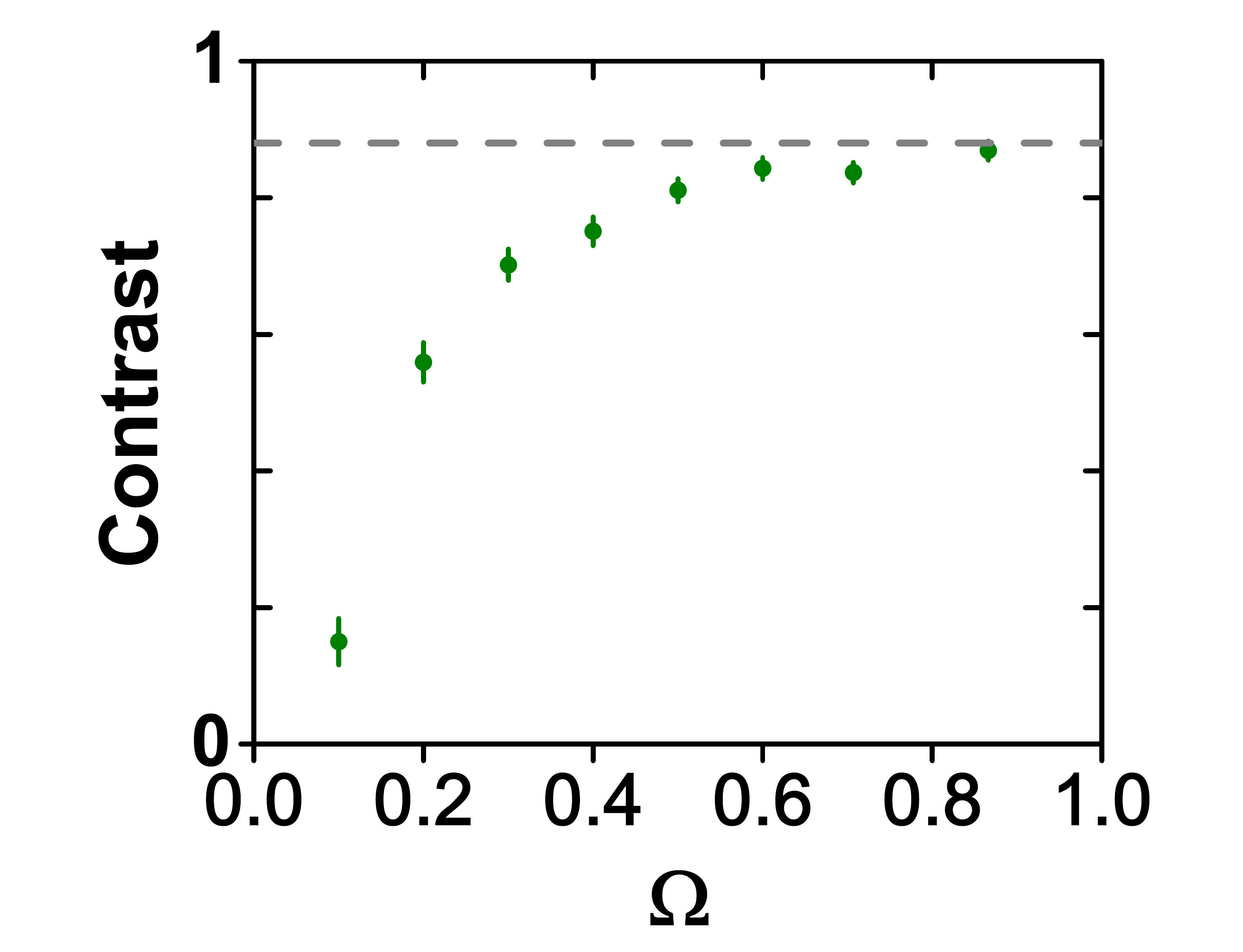}}
\caption{\label{EDfig3}
Robust microwave pulse scheme. 
(a) Spin echo fringe with N=100 $\pi$-pulses. 
(b) Contrast of the fringe in (a) as a function Rabi frequency of each $\pi$-pulse, 
.}
\end{figure}

While this pulse scheme works nearly perfectly in a set of four, its implementation in an RB sequence is interrupted when there is a $\pi/2$ pulse or a  Pauli pulse (see Supplementary Fig. 1). We have empirically determined the following rules for choosing the right phase when the "phase flow" is interrupted: 
\begin{enumerate}
\item The spin echo $\pi$ pulses around successive Pauli pulses (see Supplementary Fig. 1) should follow the cycle : \{$y,-y,-y,y$\}. 
\item If the PG preceding a CG has a torque vector along the same axis as the CG's $\pi/2$ pulse, then the torque vector of the first $\pi$ pulse inside the CG should be opposite that of PG. Otherwise, it can have the same phase as the $\pi/2$ pulse.
\item The final four $\pi$ pulses in a CG should follow the cycle : \{$y,y,-y,-y$\}.     
\end{enumerate}
These rules are based on experimental testing with many randomized benchmarking sequences and phase schemes, but they may not be optimal. 

\begin{figure}[!t]
\centering
\subfigure[t][]{
\includegraphics[width=3.1in]{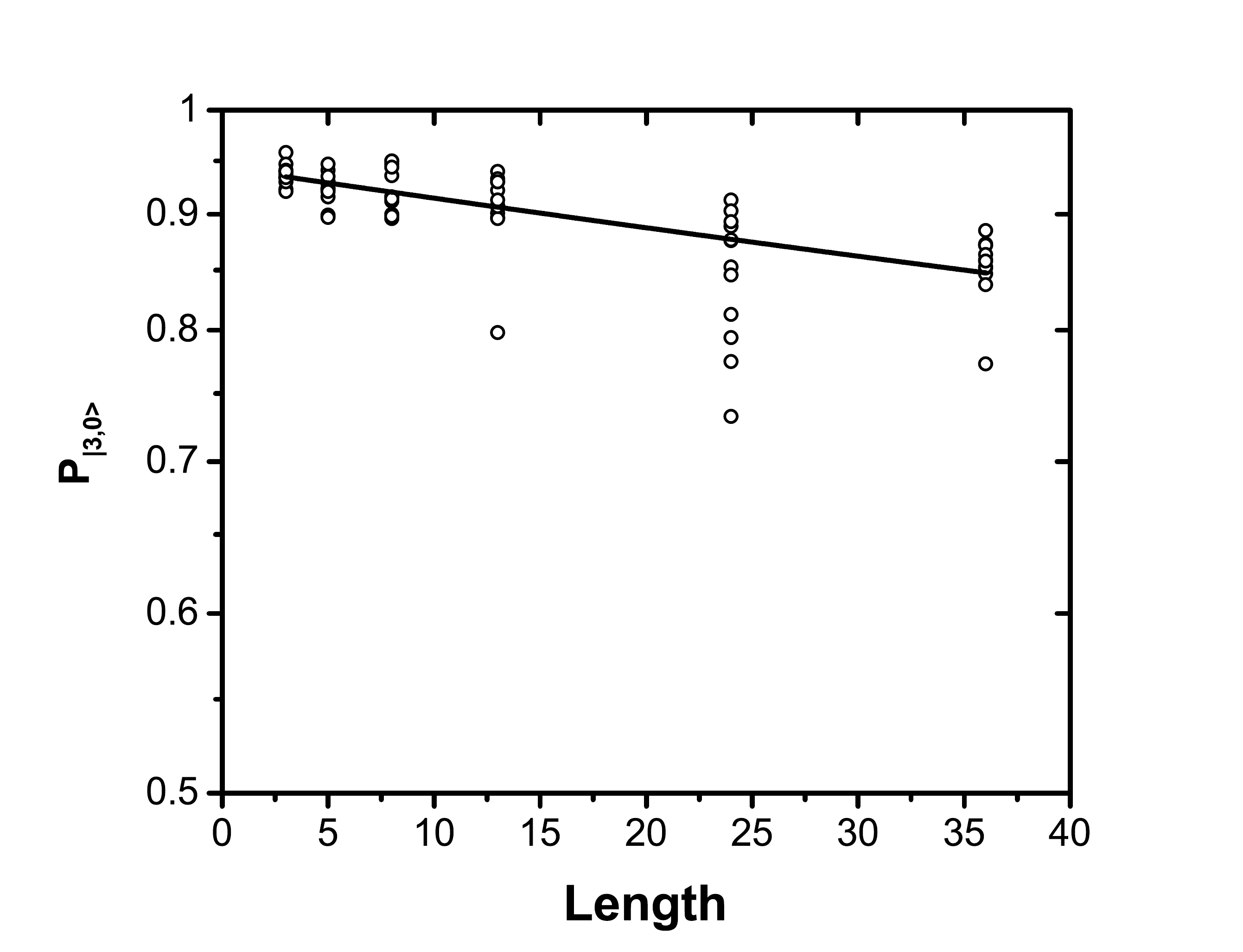}}
\subfigure[t][]{
\includegraphics[width = 3.1in]{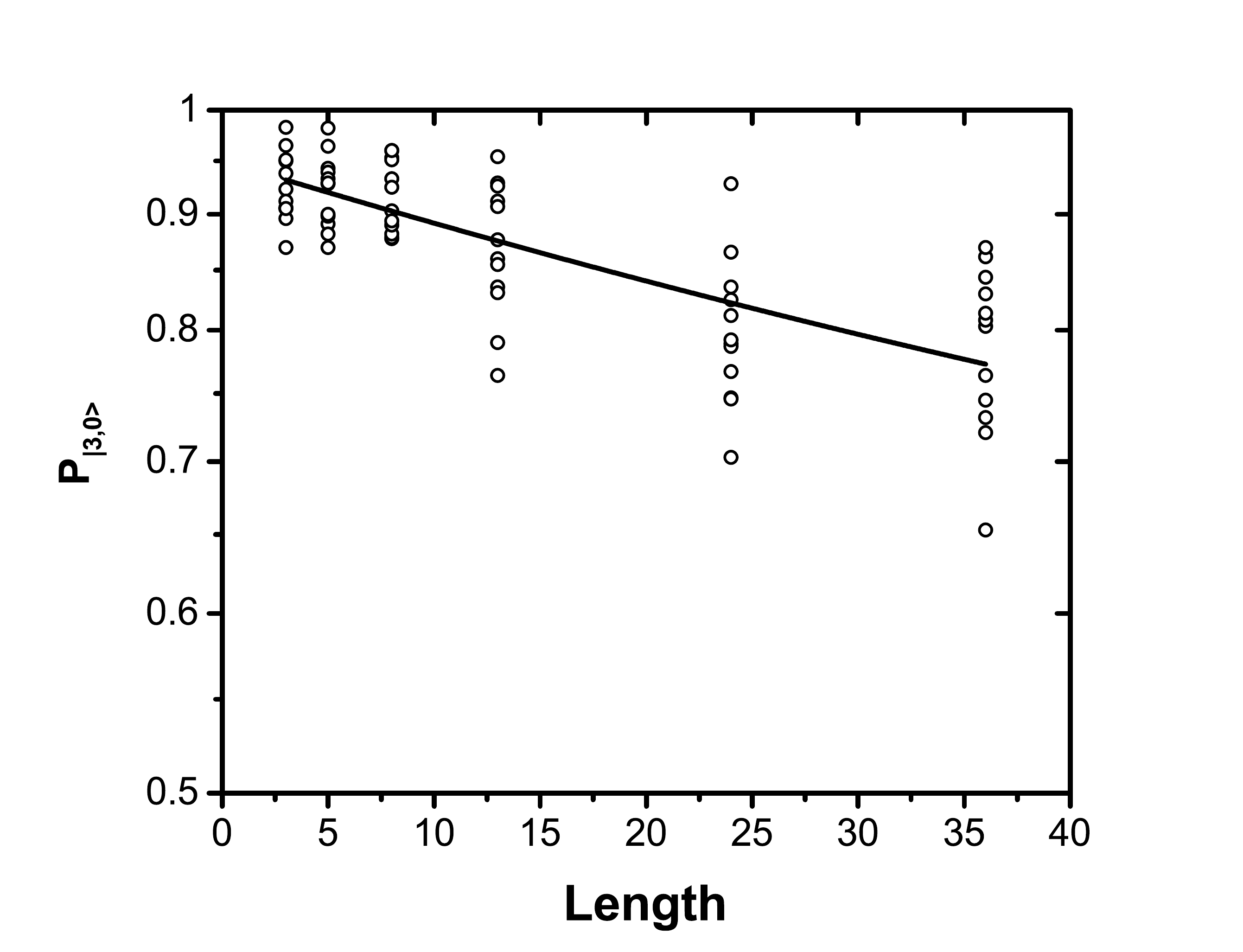}}
\caption{\label{EDfig2}
RB on non-target atoms. For each length, 3 randomized CG sequence, each of which combined with 4 sets of randomized PG sequence, are applied, totaling 12 points. (a) RB data for spectator atoms. (b) RB data on line atoms.}
\end{figure}

\section{Auxiliary fidelity calculation}
The fidelity for data in Fig. 2. is calculated as follows. The qubit state takes the form of $|\psi \rangle= $ $n cos \left( \frac{\theta}{2}\right) |0\rangle+ n e^ { i \phi} sin \left( \frac{\theta}{2} \right) |1\rangle $,
where $\theta$ and $\phi$ are the polar and azimuth angles on a Bloch sphere, and $n$ represents a spherically symmetric shrinkage of the Bloch sphere. We measure the population in $|1\rangle$. After a detection $\pi/ 2$ pulse with a phase $\alpha$ scanned, the population has a sinusoidal dependence on $\alpha$:
$ P_{|1\rangle} = n^2 (1+sin(\theta) cos (\alpha + \phi)) / 2$.
We first normalize the results by the maximum/minimum population (0.95/0.01), which is set by the SPAM error, and then fit the curve to the above functional form, enabling us to reconstruct the qubit state after the gate operation. 

The fidelity in the Fig. 4. inset is determined as follows.  We fix all the qubit state parameters to the theoretically perfect values ($\theta=\frac{\pi}{2}$, $n=1$), except for $\phi$, which is determined by the $R_z(\pi /2)$ gate. Since both the fidelity $\mathcal{F}^2$ and the population in $|1\rangle$ are a function of $\phi$, a relation between $\mathcal{F}^2$ and $P_{|1\rangle}$ can be established: $\mathcal{F}^2=\sqrt {P_{|1\rangle}}$. We normalize $P_{|1\rangle}$ as before, and then calculate $\mathcal{F}^2$ and its associated uncertainties. This approach is less mathematically rigorous than RB, but it is sufficient to illustrate the insensitivity of the gate to addressing beam alignment and intensity noise.

\clearpage

\end{document}